%% file: extended-abstract.tex
\documentclass{sigchi-ext}
\usepackage[T1]{fontenc}
\usepackage{textcomp}
\usepackage[scaled=.92]{helvet} 
\usepackage{graphicx} 
\usepackage{balance}  
\usepackage{booktabs} 
\usepackage{ccicons}  
\usepackage{ragged2e} 
\usepackage{amsmath,amssymb}
\usepackage{fontawesome}

\newcommand{\nb}[3]{
  \fcolorbox{black}{#2}{\bfseries\sffamily\scriptsize#1}
    {\sf\small$\blacktriangleright$\textit{#3}$\blacktriangleleft$}
}
\definecolor{lavendermist}{rgb}{0.9, 0.9, 0.98}
\definecolor{aliceblue}{rgb}{0.94, 0.97, 1.0}
\definecolor{antiquewhite}{rgb}{0.94, 1, 1}
\newcommand\agathe[1]{\nb{Agathe}{green}{#1}}
\newcommand\fanny[1]{\nb{Fanny}{blue}{#1}}

\def\plaintitle{Across the LLM Supply Chain: a Study of Stakeholders' Perceptions \& Needs} 
\def\emptyauthor{}
\def\plainkeywords{transparency, explainability, machine learning, stakeholder diversity, empirical study}

\title{
Understanding Stakeholders' Perceptions and Needs Across the LLM Supply Chain}

\numberofauthors{5}
\author{%
  \alignauthor{%
    \textbf{Agathe Balayn}\\
    \affaddr{ServiceNow, Delft University of Technology, Trento University} \\
    \email{a.m.a.balayn@tudelft.nl} }
    \alignauthor{%
    \textbf{Lorenzo Corti}\\
    \affaddr{Delft University of Technology}\\
    \email{l.corti@tudelft.nl} }
    \vfil 
    \alignauthor{%
    \textbf{Fanny Rancourt}\\
    \affaddr{ServiceNow}\\
    \email{fanny.rancourt@servicenow.com} }
    \alignauthor{%
    \textbf{Fabio Casati}\\
    \affaddr{ServiceNow}\\
    \email{fabio.casati@servicenow.com} } \vfil 
    \alignauthor{%
    \textbf{Ujwal Gadiraju}\\
    \affaddr{Delft University of Technology}\\
    \email{u.k.gadiraju@tudelft.nl} } 
    }

\definecolor{linkColor}{RGB}{6,125,233}
\hypersetup{%
  pdftitle={\plaintitle},
  pdfauthor={\emptyauthor},
  pdfkeywords={\plainkeywords},
  bookmarksnumbered,
  pdfstartview={FitH},
  colorlinks,
  citecolor=black,
  filecolor=black,
  linkcolor=black,
  urlcolor=linkColor,
  breaklinks=true,
}

\begin{document}

\CopyrightYear{2024}
\setcopyright{rightsretained}
\conferenceinfo{CHI'24, HCXAI workshop}{Honolulu, HI, USA}
\isbn{}
\doi{}
\copyrightinfo{\acmcopyright}

\maketitle

\RaggedRight{} 

\begin{abstract}
Explainability and transparency of AI systems are undeniably important, leading to several research studies and tools addressing them. Existing works fall short of accounting for the diverse stakeholders of the AI supply chain who may differ in their needs and consideration of the facets of explainability and transparency. In this paper, we argue for the need to revisit the inquiries of these vital constructs in the context of LLMs. To this end, we report on a qualitative study with 71 different stakeholders, where we explore the prevalent perceptions and needs around these concepts. This study not only confirms the importance of exploring the ``who'' in XAI and transparency for LLMs, but also reflects on best practices to do so while surfacing the often forgotten stakeholders and their information needs. Our insights suggest that researchers and practitioners should simultaneously clarify the ``who'' in considerations of explainability and transparency, the ``what'' in the information needs, and ``why'' they are needed to ensure responsible design and development across the LLM supply chain. 
\end{abstract}

\input{content}

\balance{} 

\bibliographystyle{SIGCHI-Reference-Format}
\bibliography{bibli}

\end{document}

%% file: content.tex
\section{A Case for XAI Across the Supply Chain}


Research addressing explainability and transparency of AI systems seldom clarifies the purpose and the target stakeholders, 
often assuming the information needs of stakeholders with little or no validation \cite{mitchell2019model,gebru2021datasheets,smilkov2017smoothgrad}.
There has been an increasing recognition that the development of AI systems 
spans a long supply chain of stakeholders \cite{widder2023dislocated,cobbe2023understanding,piorkowski2022evaluating}. However, only AI developers \cite{balayn2022can,delaunay2023adaptation} and end-users \cite{kim2023help,ehsan2021expanding} have received considerable attention thus far. Other stakeholders such as legal teams or product managers remain largely absent in existing explorations.
\marginpar{%
   \vspace{-50pt} 
  \fbox{%
    \begin{minipage}{0.925\marginparwidth}
      \textbf{S1 -- Study participants. } 
      We cover a plurality of domains in the supply chain, including engineering (AI developers, AI researchers), user experience (user research, content design), business (product managers, business analysts, customer service), and governance (legal and risk teams, government relations). 
      The practical tasks of our participants involve, e.g., implementing, testing, and overseeing AI systems, 
      or strategizing around the requirements of the system and broader organization (e.g., functional requirements, definition of risks, etc.).
      These participants' responsibilities span executives, managers, less senior employees, etc.
      Finally, we also interviewed end-users of the AI systems developed by our participants --they are internal to the companies developing the AI systems or independent thereof, and none are AI experts.
    \end{minipage}}\label{sidebar:example} }
Inspired by prior work \cite{dhanorkar2021needs,ehsan2021explainable,heger2022understanding,yurrita2023generating}, we argue for the necessity to broaden our understanding of AI stakeholders, their engagement and needs pertaining to explanations and transparency. 

Particularly, in an era in which AI systems are increasingly closed-source and concealed behind API access (e.g., LLM-based chatbots), stakeholders may elicit 
\textit{new} 
needs for explainability and transparency. 
Following is an example of such information needs across stakeholders.
An \textit{applied researcher} within organization A might be looking for an open-source foundation model of any organization and might select the model of \textit{organization B} that is pre-trained on the dataset containing the most relevant data samples to the application they are developing. Then, the \textit{legal team} of organization A might have to enquire about the \textit{data collection practices} of organization B, e.g., to make sure that they respect licensing agreements \cite{paullada2021data} and labor rights of data annotators \cite{altenried2020platform}. Finally, the \textit{quality team} of organization C deploying the AI system they would buy from organization A would need to inspect explanations about the outputs of the system to make sure that the system performs well-enough for their application.

It is complex and challenging, yet vital to investigate and 
surface such underexplored aspects in the AI supply chain and relevant 
practices across organizations. 
Aligned with the notion of interpretative flexibility \cite{meyer2006three}, stakeholders' perceptions of AI systems directly affect their needs and collaborative practices across the AI supply chain. It can therefore be challenging to appropriately investigate known and unknown explainability and transparency needs. 
\\
In this work, we take a formative step towards characterising ``who'' the stakeholders of explainability and transparency are across the AI supply chain and ``what'' their information needs are. 
We report early results from a qualitative inquiry about the perceptions and needs for explainability and transparency 
of 71 AI stakeholders across different organizations. 
Our results underscore the importance of grounding studies in the \textit{who}, \textit{what}, and \textit{why} of explainability and transparency, reflect on methodological best practices, 
and bear implications for the HCXAI community in establishing and shaping relevant policies.

\section{Study Methodology}

We interviewed 71 participants, across 10 private organizations, and spanning a wide variety of roles (see \textbf{S1}). Participants were recruited via our professional network, with snowball and convenience sampling, to recruit a diversity of stakeholders. All participants develop or use LLMs.
In the interviews (on average 45 mins.), we first asked our participants to describe their day-to-day activities, and to discuss their primary concerns, challenges, and needs concerning the harmful impact of AI systems. This allowed us to 
identify information needs. 
We then probed their understanding of explainability and transparency: what these concepts evoke, how important they are, and what the entailing challenges might be. 
We first analysed the interviews through deductive coding looking for the ``what'', ``who'', and challenges of the informational needs across stakeholders. Then, we performed a round of inductive coding to identify other recurrent themes. 

\section{Results}

\textbf{1) The ``\textit{what}": Explainability and Transparency} \\
Generally, our participants viewed transparency as encompassing information beyond AI models (whereas XAI was viewed as being concerned with these models) such as training data, code, and reports about other supply chain stakeholders.
Some information needs related to the outputs of the models, be it their trustworthiness (accuracy or fairness), surrounding information  (e.g., confidence scores), or explanations (as understood by the algorithmic XAI community). Other information items related to the internal workings of the LLM (e.g., details about its training dataset) or its 'meta-properties' (e.g., organization developing it). 

Our study points toward several information needs that have been discussed in prior literature \cite{mitchell2019model,gebru2021datasheets,anik2021data,bommasani2023foundation}.
Interestingly enough, more recent artefacts such as the Foundation Model Transparency Index \cite{bommasani2023foundation}, cover elements that our participants perceived as general concerns rather than specific to explainability and transparency (see \textbf{S2}). These include data labor, environmental impact, and broader ethical concerns.
Regardless, participants focused on elements that are often considered incidental for XAI and transparency (e.g., confidence values for individual predictions) or ``less-technical’’ (e.g., developers' discretionary choices, discussed by a few prior works~\cite{ehsan2021expanding,piorkowski2022evaluating}). 
Some participants perceived XAI and transparency as information primarily useful for end-users and AI system deployers. Disclosure on LLM-generated samples (18\% of participants) and basic explanations pertaining to LLMs (24\% of participants) were of particular interest.  Other participants discussed their own information needs. 

\marginpar{%
  \vspace{-300pt} 
  \fbox{%
    \begin{minipage}{0.925\marginparwidth}
      \textbf{S2 -- Information items discussed by our AI stakeholders. }

     \footnotesize
     \begin{tabular}{p{3.2cm}p{0.3cm}}

\toprule
Information need  & \% \\
\midrule
\textbf{"LLM performance":} \\
 fairness-metric related & 2 \\
limits of LLM & 22 \\
\textbf{"LLM decisions":}\\
confidence for each LLM output & 2 \\
 fact that output is LLM-generated & 18 \\
 log of LLM inputs \& outputs & 4 \\
traceability of errors \& outputs & 9 \\
\textbf{Explainability:}\\
explanations of LLM output via LLM input \& addit. info used by LLM & 20 \\
 LLM system rationales leading to an LLM output & 47 \\
\textbf{"LLM internals":}\\
LLM design decisions & 53 \\
 LLM evaluation decisions & 18 \\
 customer data handling & 20 \\
\textbf{"Meta info":}\\
basic explanations about LLM & 24 \\
 guidelines for good use of LLM  by customer & 9 \\
 who are the creators & 2 \\
open source  & 2 \\
what other teams do about LLM or for a specific LLM & 4 \\

\end{tabular}
    \end{minipage}
    }\label{sidebar:example} 
    }

\begin{figure}[h!]
    \centering
    \includegraphics[width=0.95\columnwidth]{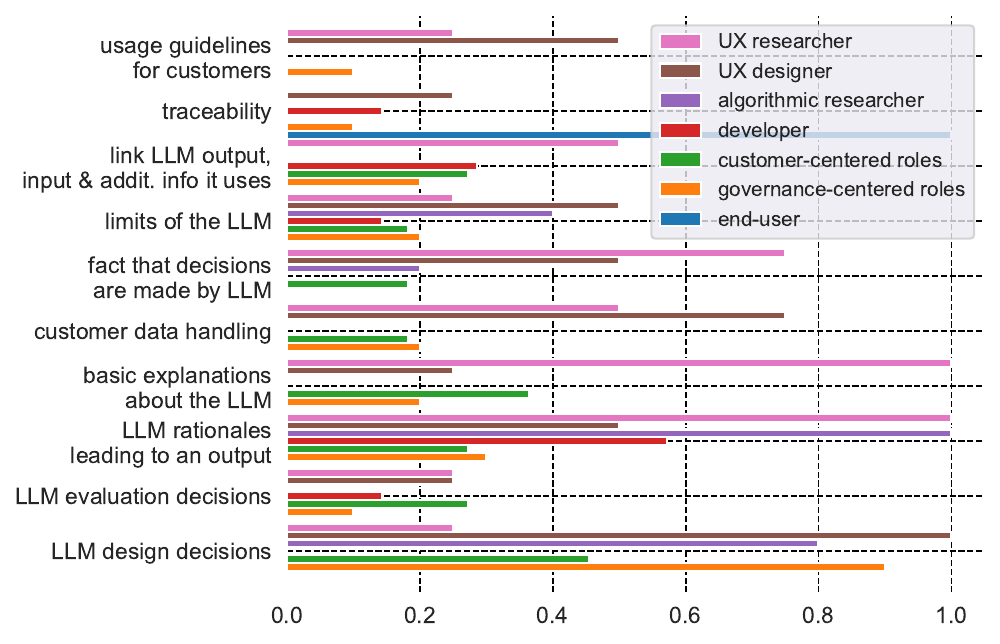}
    \caption{Information needs per category of stakeholders (fraction of the participants who mentioned the need). We excluded information needs mentioned by a single category of stakeholders:
     developers - LLM confidence (14\%); customer-centered roles - log of LLM inputs/outputs (18\%), fairness performance (9\%); UX researcher - open-source (25\%), work of other teams (50\%); end-user - LLM creators (100\%).}
    \label{fig:enter-label}
\end{figure}

\textbf{2) The ``why": Connecting ``what'' and ``who''} 
\\Given the broad range of AI stakeholders interviewed, we note several trends concerning the purposes for explainability and transparency. Unsurprisingly, such trends are shaped by a stakeholder’s role, prior knowledge, and network of collaborators (see Figure \ref{fig:enter-label}).
Developers, algorithmic researchers, and quality assessors frequently mentioned technical information (e.g., prediction confidence and design decisions) necessary to select which AI model to fine-tune, debug the overall system, or identify the potential algorithmic harms.
On the other hand, UX designers and researchers provide guidelines to customers for responsible use of the AI system. Together with product managers, they emphasized disclosing essential information to customers or end-users (e.g., to gain buy-in) or to ethics review boards to attest to Responsible AI endeavors (e.g., appropriate handling of customer data).
Such a purpose appears to be met by end-users’ need for transparency from the makers of AI-powered products (e.g., to gauge the trustworthiness of a company).
Finally, we note that XAI and transparency were attributed cross-functional purposes around fostering  Responsible AI practices and accountability internally and externally across the AI supply chain.

\textbf{3) Challenges faced by AI stakeholders}
\\Our study highlighted several challenges AI stakeholders face concerning explainability and transparency both at an individual and collaborative level.

At an individual level, participants expressed feasibility tensions in implementing explainability and transparency, e.g., between XAI and customers’ data confidentiality. 
For instance, the quality assessment team discussed challenges with \textit{information access}. They reported on their need to get to know customers and end-users of the LLMs, to appropriately assess these LLMs (e.g., considering social biases). Yet, they emphasized that accessing any information about such external stakeholders is impossible from their position in the AI supply chain.
Additionally, technical implementation details are often difficult to accurately report due to (1) the fast-paced development of LLMs, (2) potential exposure to adversarial attacks, and (3) business secrets. 

Participants in our study reported mixed needs for disclosure or concealment of information.
Broadly, at a collaborative level, many stakeholders appear to \textit{lack knowledge} about the functioning of AI and explainability and transparency.
For instance, several deployers of AI systems recognized not being aware of potential harms caused by AI systems and hence not being able to ask for relevant information in that regard. Other participants, such as some algorithmic researchers, were only cognizant of algorithmic XAI concepts and debates around AI open-source, which shows a \textit{narrow perspective} on XAI.
This prevents them from identifying and understanding information relevant to either their or others' work.

As a consequence, we identified \textit{misconceptions} (e.g., a participant argued that making an AI system transparent necessarily leads to ensuring the fairness of its outputs) and \textit{terminological confusion} around XAI and transparency (e.g., most participants use "transparency", "explainability" or "accountability" assuming a certain meaning that they disentangle only when prompted), which, more broadly, can cause miscommunication and mishandling of AI systems.
Finally, participants’ surroundings and time constraints (owing to the ongoing LLMs race) affect their views on XAI and transparency, hampering appropriate reflections.

\section{Discussion \& Implications}

Our results 
emphasise the importance of adopting a broad view of the AI supply chain when discussing XAI and transparency, illustrating an existing lack of understanding of explainability and transparency across it. As prior works (e.g., \cite{bhatt2020explainable,anik2021data,heger2022understanding,norkute2021ai,preece2018stakeholders,langer2021we}) had not identified as many stakeholders and their purposes \cite{suresh2021beyond,sokol2020explainability,brennen2020people}, nor all information items, we argue that there is a strong need for future work to be more comprehensive. 
The supply chain view also enabled us to surface concrete factors that impact information needs, and diverse, complex challenges to be tackled in the future. 
Our preliminary insights provide a range of methodological lessons for future work. \\
%
%

%
\begin{itemize}[leftmargin=*]
    \item The complex relations between the ``what", ``who", and ``why" 
    echo the need to investigate XAI and transparency in context, without solely focusing on the information items (i.e., the ``what"). This reaffirms a call for  future work on XAI and transparency to be grounded within the ``who" and ``why", as proposed, e.g., by the AI FactSheet methodology \cite{richards2020methodology,piorkowski2022evaluating}. While many XAI and transparency tools have been proposed (e.g., \cite{gebru2021datasheets,mitchell2019model,bommasani2023foundation,anik2021data}), they focus solely on the ``what" and require revisiting 
    in light of our stakeholders' needs  and concomitant challenges.
    \item Our study emphasizes the need for methodological triangulation 
    to develop a holistic understanding of stakeholders' perceptions and needs across the AI supply chain. 
    XAI and transparency are complex constructs with a nuanced and overlapping scope. They are perceived differently by those in different positions within the supply chain, present great terminological ambiguity (not to be mistaken with diversity), 
    and evoke different information items, stakeholders, and purposes. 
    In a sense, they are boundary objects to discuss information exchanges and needs across stakeholders. 
    In our study, we used these constructs as starting points for discussions. 
    We observed that 
    AI stakeholders referred to different information needs for broader goals while discussing their current work and concerns. 
    E.g., knowledge of data access rights per customer was discussed by AI developers for training or evaluating the AI system but not in the context of XAI and transparency. 
    Hence, we believe that future studies would benefit from adopting complementary objects of study: studying not only perceptions of XAI, transparency and the needs thereof,  
    but also in-depth information needs of the different stakeholders while refraining from referring to the terms XAI and transparency.
    \item We recommend bearing in mind stakeholders' different interpretations of AI ``explainability'' and ``transparency'' --often confused-- and ensuring conceptual clarity in discussions involving these terms. This can be done by asking participants to clarify their use of the terms or providing clear definitions 
    depending on the goal of the studies. 
    \item Our findings can aid the development of a conceptual framework around AI information needs, that could be used to carefully craft further studies. 
    Such a framework could start with the identified who, what, and why of AI information needs, and also encompass personal obstacles, organizational challenges, tensions and preferences. 
\end{itemize}

\textit{\textbf{Policy, Explainability, and Transparency.}}\\
Our results finally point out the impossibility of solely relying on stakeholders' perceived information needs to define  XAI and transparency obligations. 
Indeed, this topic was revealed to be highly dependent on stakeholders' knowledge and on their subjective judgment of the various tensions highlighted -- many information items found in prior work were not mentioned by any participant, with or without a technical role or background, despite all of them being decision makers across the supply chain and about explainability and transparency considerations. The identified informational challenges hint at the potential undesirability to answer any information need.
Hence, ``who" should be responsible for making such decisions remains to be explored both within organizations and on a regulatory level.